\documentclass[aps,prd,amsmath,twocolumn,amssymbaps,showpacs]{revtex4-1}
\usepackage{graphicx}  
\usepackage{dcolumn}   
\usepackage{bm}        
\usepackage{amssymb}   
\usepackage{amsmath}   
\usepackage{lineno}
\usepackage{draftcopy}
\usepackage{relsize} 
\usepackage{xfrac}    

\usepackage{color}
\definecolor{dgreen}{cmyk}{1.,0.,1.,0.2}        
\definecolor{orange}{cmyk}{0.,0.353,1.,0.}    

\usepackage[bookmarks]{hyperref}

\newcommand{\di}{{\rm d}}

\newcommand{\be}{\begin{equation}}
\newcommand{\ee}{\end{equation}}                                                                               
\newcommand{\bea}{\begin{eqnarray}}
\newcommand{\eea}{\end{eqnarray}}

\begin{document}
\title{Rotation induced charged pion condensation in a strong magnetic field:\\ A Nambu--Jona-Lasino model study}

\author{Gaoqing Cao$^1$ and Lianyi He$^{2}$}
\affiliation{1 School of Physics and Astronomy, Sun Yat-Sen University, Guangzhou 510275, China\\ 2 Department of Physics, Tsinghua University, Beijing 100084, China}
\date{\today}

\begin{abstract}
We investigate the possibility of charged pion condensation in the presence of parallel rotation and magnetic field within the Nambu--Jona-Lasinio model with quarks as the fundamental degrees of freedom. 
Previous study based on non-interacting Klein-Gordon theory for pions showed that the charged pions will undergo Bose-Einstein condensation under this circumstance [Y.~Liu and I.~Zahed,
Phys.\ Rev.\ Lett.\  {\bf 120},  032001 (2018)]. In this work, we take into account the internal quark structures of charged pions self-consistently through quark polarization loops in an interacting theory, i.e., 
the Nambu--Jona-Lasino model. The stability of the system against the formation of charged pion condensation, i.e., a nonzero expectation value of the composite charged pion field $\bar{u}i\gamma_5d$, is explored. 
We find that two competing effects are induced by the rotation:  the isospin enhancement which favors charged pion condensation and the spin breaking which disfavors the condensation.  For a strong magnetic field 
($\sqrt{eB}\sim1{\rm GeV}$) and system size of a few fermi,  the isospin enhancement effect is stronger than the spin breaking one, and the charged pion condensation becomes energetically favored beyond a critical 
angular velocity of a few MeV.
\end{abstract}

\maketitle

\section{Introduction}

Relativistic heavy ion collisions are able to realize several extreme conditions ever found in our Universe: ultra-high temperature~\cite{Adare:2008ab,Wilde:2012wc},  strong electromagnetic (EM) field~\cite{Skokov:2009qp,Deng:2012pc}, and fast rotation~\cite{Liang:2004ph,STAR:2017ckg,Becattini:2016gvu}. The properties of quantum chromodynamics (QCD) at ultra-high temperature, i.e., the quark-gluon plasma, can be experimentally studied in central collisions~\cite{Adler:2004zn,Aamodt:2010cz}. On the other hand, it has been noticed in recent years that strong EM field and fast rotation can be realized in peripheral collisions where large electric charges of the spectators and angular momentum are involved~\cite{Bloczynski:2012en,Deng:2016yru,Karpenko:2016jyx,Niida:2018hfw,Guo:2019joy}. However, the properties of QCD matter at strong EM field and fast rotation are quite mysterious, including several unsolved problems such as the inverse magnetic catalysis effect~\cite{Bali:2011qj,Bali:2012zg} and the "sign puzzles" of the local polarizations of $\Lambda$ hyperon~\cite{Niida:2018hfw,Becattini:2017gcx,Xia:2018tes,Becattini:2019ntv,Xia:2019fjf}.
Moreover, at high enough temperature where the chiral symmetry is restored and the quarks become almost massless, the QCD matter shows intriguing features at strong EM field and fast rotation induced by the quantum anomaly of chiral fermions, such as the chiral magnetic effect and the chiral vortical effect ~\cite{Liao:2014ava,Kharzeev:2015znc,Huang:2015oca}.

In this work, we focus on QCD matter in a specific circumstance with parallel rotation and magnetic field (PRM), for which there are already several interesting findings. The effect of PRM on the dynamical chiral symmetry breaking was studied in~\cite{Chen:2015hfc}. It shows that the rotation plays the same role as the baryon chemical potential and thus leads to chiral symmetry restoration even in a strong magnetic field.  The transport properties of chiral fermions in PRM was investigated in~\cite{Hattori:2016njk}.  It was found that  in the lowest Landau level (LLL) approximation,  PRM would induce axial current along the direction of the magnetic field or angular velocity, i.e.,  the so called ``anomalous magnetovorticity effect". While these studies were based on theories with quarks as elementary degrees of freedom,  another exploration based on theories with mesons (or explicitly pions) as elementary degrees of freedom~\cite{Liu:2017spl} found that charged pion condensation (CPC) forms at sufficiently fast rotation, due to the energy splitting of $\pi^+$ and $\pi^-$ mesons under PRM.

Although it is very interesting to explore the interplay among the dynamical chiral symmetry breaking, the charged pion condensation, and the magnetovorticity effect under PRM, this work is devoted to study the possibility of CPC based on an interacting chiral effective model with quarks as elementary degrees of freedom, i.e., the Nambu--Jona-Lasino (NJL) model~\cite{Nambu:1961tp,Nambu:1961fr,Klevansky:1992qe}. It is known that in a strong magnetic field, even the neutral pion is strongly influenced through its internal quark structure~\cite{Bali:2017ian,Wang:2017vtn,Mao:2018dqe,Liu:2018zag,Cao:2019res}. However, previous study based on non-interacting Klein-Gordon theory of charged pions ignored their internal quark structures~\cite{Liu:2017spl}.  Therefore, it is necessary to check the possibility of CPC induced by PRM using an interacting theory with quarks as elementary degrees of freedom. Our motivation is as follows. We take $\pi^+=\bar{d}i\gamma^5u$ for example. Applying a magnetic field will align both the spins of $u$ and anti-$d$ quarks along its direction because both quarks have positive charges. If we further turn on 
a parallel rotation in the system, the situation does not change. Finally, we would expect the total spin of $u$ and anti-$d$ quarks to be $1$ because of the polarization effect, which contradicts with the pseudoscalar nature of $\pi^+$. According to this spin breaking picture, we may conclude that CPC is not favored in PRM, similar to the pure magnetic field case~\cite{Cao:2015xja}.   

The CPC was previously predicted to the ground state of QCD at finite isospin density, when the isospin chemical potential becomes larger than the vacuum pion mass~\cite{Son2001,Cohen:2003kd,Barducci2004,Loewe2004,He2005,He2006,Ebert2006,Zhang2007,Andersen2007,Carignano2017,Brandt2018}. With increasing isospin density, the system undergoes a crossover from a Bose-Einstein condensation of charged pions to a Bardeen-Cooper-Schrieffer superfluid of Cooper pairs of quarks and anti-quarks~\cite{He2006}. The effect of a magnetic field on the CPC at finite isospin density was also studied within the NJL model~\cite{Cao:2015xja}, Lattice QCD simulation~\cite{Endrodi:2014lja}, linear sigma model~\cite{Loewe:2013coa}, and Klein-Gordon theory~\cite{Ayala:2016awt}. It was shown that the critical temperature of CPC increases with increasing magnetic field at a given isospin density~\cite{Loewe:2013coa,Ayala:2016awt}.  Besides, the chiral soliton lattice,  an inhomogeneous state of neutral pion condensation was predicted in the chiral perturbation theory at finite baryon chemical potential with magnetic field~\cite{Brauner:2016pko}. However, it was shown to be unstable to CPC at large magnetic field or baryon chemical potential~\cite{Brauner:2016pko}. 

In this work, we focus on the possibility of CPC induced by rotation in a strong magnetic field. In addition to the spin breaking effect in a magnetic field, the rotation will play the role of an effective isospin chemical potential, which favors the CPC. We study the stability of the system against the formation of  CPC, i.e., a nonzero expectation value of the composite charged pion field $\bar{u}i\gamma_5d$. We show that in the presence of PRM, there exist  two competing effects: the isospin enhancement which favors charged pion condensation and the spin breaking which disfavors the condensation.  For a strong magnetic field ($\sqrt{eB}\sim1{\rm GeV}$) and system size of a few fermi,  the isospin enhancement effect is stronger than the spin breaking one. We find that the charged pion condensation becomes energetically favored beyond a critical angular velocity of a few MeV, much smaller than the energy scale of the magnetic field.

The paper is organized as follows. In Sec.\ref{model}, we present the NJL model in rotating frame with a parallel magnetic field, construct the quark propagator and then obtain the gap equation for the dynamical quark mass. In Sec.\ref{numerical}, the stability of the QCD system against CPC is studied via a Ginzburg-Landau-like approach and the numerical results are illuminated. We summarize in Sec.\ref{conclusions}. The natural units $c=\hbar=k_{\rm B}=1$ is used throughout.

\section{Nambu--Jona-Lasinio model in rotating frame}\label{model}

In order to explore the influence of the internal structures of pions on the formation of charged pion condensation, we adopt the two-flavor NJL model with $u$ and $d$ quarks as the fundamental degrees of freedom~\cite{Klevansky:1992qe}. We consider a system under the circumstance of a global rotation with  angular velocity $\boldsymbol{\Omega}=\Omega\hat{z}$ and a constant parallel magnetic field ${\bf B}=B\hat{z}$.  In the rotating frame, the action of the system can be conveniently given in curved spacetime by 
\begin{eqnarray}
{\cal S}=\int d^4x \sqrt{-\det(g_{\mu\nu})}{\cal L}(\bar{\psi},\psi),
\end{eqnarray}
where the Lagrangian density is given by
\begin{eqnarray}\label{Lcv}
{\cal L}&=&\bar\psi\left[i\gamma^\mu(D_\mu\!+\!\Gamma_\mu)\!-\!m_0\right]\psi\!+{\cal L}_{\rm int}\nonumber\\
{\cal L}_{\rm int}&=&\!G\left[(\bar\psi\psi)^2\!+\!(\bar\psi i\gamma_5\mbox{\boldmath{$\tau$}}\psi)^2\right].
\end{eqnarray}
Here $\psi=(u,d)^{\rm T}$ represents the two-flavor quark field, $m_0$ is the current quark mass, $G$ is the coupling constant, $\tau_i$ ($i=1,2,3$) are the Pauli matrices in the flavor space, and
\begin{eqnarray}
D_\mu=\partial_\mu+i\hat{q}A_\mu
\end{eqnarray}
is the covariant derivative, with the electric charge matrix $\hat{q}={\rm diag}(q_u,q_d)$ and the vector potential $A_\mu$ presenting EM field in curved spacetime. 

The rotating frame is characterized by the spacetime metric given by 
\begin{eqnarray}\label{gcv}
g_{\mu\nu}=\left(\begin{array}{cccc}
1-(x^2+y^2)\Omega^2&y\Omega&-x\Omega&0\\
y\Omega&-1&0&0\\
-x\Omega&0&-1&0\\
0&0&0&-1
\end{array}\right),
\end{eqnarray}
then $\det(g_{\mu\nu})=-1$. The coupling to spin is presented by the affine connection $\Gamma_\mu$ which is defined in terms of the spin connection $\omega_{\mu ij}$ and the vierbein $e_{i}^\mu$ as
\begin{eqnarray}\label{gammacv}
&&\Gamma_\mu=-{i\over4}\omega_{\mu ij}\sigma^{ij}, \nonumber\\
&&\omega_{\mu ij}=g_{\alpha\beta}e^\alpha_i\left(\partial_\mu e^\beta_j+\Gamma^\beta_{\mu\nu}e^\nu_j\right),\nonumber\\
&&\sigma^{ij}=\frac{i}{2}\left[\gamma^i,\gamma^j\right].
\end{eqnarray}
The Greek and Latin letters denote the indices in
coordinate and tangent spaces, respectively, and $\Gamma^\beta_{\mu\nu}$ is the Christoffel symbol related to the derivatives of the metric $g_{\mu\nu}$ with respect to the coordinates. In the following, we can choose $e_0^t=e_1^x=e_2^y=e_3^z=1$, $e_0^x=y\Omega$, $e_0^y=-x\Omega$, and other components are zero~\cite{Jiang:2016wvv}. 

The symmetric gauge will be chosen for the vector potential in the inertial frame, that is
$A_i=(0,By/2,-Bx/2,0)$, which just results in ${\bf B}=B\hat{z}$. By substituting Eq.(\ref{gcv}) and (\ref{gammacv}) into Eq.(\ref{Lcv}), we eventually obtain~\cite{Chen:2015hfc}:
	\begin{eqnarray}\label{NJL}
	{\cal L}&=&\bar\psi\Big\{\gamma^0\left[i\partial_t+\Omega \left(\hat{L}_z+\hat{S}_z\right)\right]+i\gamma^1D_x+i\gamma^2D_y \nonumber\\
	&&+\ i\gamma^3\partial_z-m_0\Big\}\psi+{\cal L}_{\rm int},
	\end{eqnarray}
with the orbital angular momentum operator $\hat{L}_z\equiv-i(x\partial_y-y\partial_x)$ and the spin operator $\hat{S}_z\equiv \sigma^{12}/2$. We have defined $D_x=\partial_x+i\hat{q}By/2$ and $D_y=\partial_y-i\hat{q}Bx/2$
for convenience. Note that in the rotating frame, the vector potential is given by $A_\mu=A_ie_\mu^i=(-B\Omega r^2/2,By/2,-Bx/2,0)$, which leads to an electric field. However,  $A_0=-B\Omega r^2/2$ does not appear in Eq.(\ref{NJL}) because the gamma matrix $\gamma^0=\gamma^i e_i^t$ cancels it out~\cite{Chen:2015hfc}.

\subsection{Fermion propagator}\label{propagator}
We first derive the fermion propagator under PRM background. To this end, we consider the one-flavor fermion with constant mass $m$ and electric charge $q$ for convenience. The Dirac equation is given by
	\begin{eqnarray}\label{L1f}
	&&\Big\{\gamma^0\left[i\partial_t+\Omega \left(\hat{L}_z+\hat{S}_z\right)\right]+i\gamma^1D_x+i\gamma^2D_y \nonumber\\
	&&\ \ +\ i\gamma^3\partial_z-m\Big\}\psi=0.
	\end{eqnarray}
	
Consider the case $qB>0$ first. Working in cylindrical coordinate system with $x=r\cos\theta$ and $y=r\sin\theta$, the solution of this Dirac equation with energy $E$ can be expressed as~\cite{Chen:2015hfc}
\begin{eqnarray}
&&\psi=e^{-i\, (E t-p_zz)}H_{n,l}(\theta,r)~u_{n,l}(p_z),\nonumber\\
&&H_{n,l}(\theta,r)={\cal P}_\uparrow\chi_{n,l}(\theta,r)+{\cal P}_\downarrow\chi_{n-1,l+1}(\theta,r),
\end{eqnarray}
where ${\cal P}_{\uparrow,\downarrow}={1\over2}(1\pm\sigma^{12})$ are the spin projectors, and the normalized wave function $\chi_{n,l}$ is given by
\begin{eqnarray}
\chi_{n,l}(\theta,r)=\left[{qB\over2\pi}{ n!\over(n+l)!}\right]^{1\over2}{e^{i\, l\theta}}~\tilde{r}^le^{-\tilde{r}^2/2}L_n^l\left(\tilde{r}^2\right)
\end{eqnarray}
with $\tilde{r}^2=|qB|r^2/2$. The Laguerre polynomial $L_n^l(x)$ is nonvanishing only for $n\ge0$ and $l\in\left[-n,N-n\right]$, where the degeneracy factor $N$ for each Landau level reads
\begin{eqnarray}
N=\left \lfloor \frac{qBS}{2\pi} \right \rfloor,
\end{eqnarray}
with $S$ being the area of the $xy$-plane. Using the identity 
\begin{equation}
i\gamma^1D_x+i\gamma^2D_y=\left[{\cal P}_\downarrow(D_x+iD_y)-{\cal P}_\uparrow(D_x-iD_y)\right]\gamma^2
\end{equation}
and the relations
\begin{eqnarray}
\partial_x=\cos\theta\,\partial_r-{\sin\theta\over r}\,\partial_\theta,\ \
\partial_y=\sin\theta\,\partial_r+{\cos\theta\over r}\,\partial_\theta,
\end{eqnarray}
we obtain
\begin{eqnarray}
\left(i\gamma^1D_x+i\gamma^2D_y\right)H_{n,l}(\theta,r)=-H_{n,l}(\theta,r)\gamma^2\sqrt{2n qB}.\nonumber\\
\end{eqnarray}
Substituting this result into the Dirac equation (\ref{L1f}), we obtain the eigenvalue equation for $u_{n,l}(p_z)$:
\begin{eqnarray}
\left(\gamma^0\varepsilon^+-\gamma^3p_z-\gamma^2\sqrt{2n qB}-m\right)u_{n,l}(p_z)=0,
\end{eqnarray}
with $\varepsilon^+\equiv E+\Omega \left(l+{1\over2}\right)$. One branch of solutions, corresponding to the positive-energy ones in the absence of magnetic field and rotation, can be expressed in a compact form as~\cite{Peskin1995}
\begin{eqnarray}
u_{n,l}^s(p_z)=\left(\begin{array}{c}
\sqrt{p\cdot{\sigma}}\xi^s\\\sqrt{p\cdot\bar{\sigma}}\xi^s\end{array}\right),   
\end{eqnarray}
where $s=\pm$, $p_\mu=(\varepsilon_n,0,\sqrt{2n qB},p_z)$ with $\varepsilon_n\equiv(2n qB+p_z^2+m^2)^{1/2}$, $\sigma^\mu=(1,\mbox{\boldmath{$\sigma$}})$, and $\bar{\sigma}^\mu=(1,-\mbox{\boldmath{$\sigma$}})$. The two-component spinors $\xi^\pm$ are given by 
$\xi^+=(1,0)^{\rm T}$ and $\xi^-=(0,1)^{\rm T}$.

With the solutions of the Dirac equation, we can construct the fermion Green's function by following Ref.~\cite{Cao:2014uva}. In the following, we use the notation $x= (t,r,\theta,z)$ for convenience. For $t>t^\prime$, the retarded Green's function can be evaluated as
\begin{widetext}
\begin{eqnarray}
S_{\rm R}(x,x^\prime)&\equiv&\Theta(t-t^\prime)\langle0|\{\psi(x)\bar{\psi}(x^\prime)\}|0\rangle\nonumber\\
&=&\!\sum_{n=0}^\infty\!\sum_l\!\!\!\int_{-\infty}^{\infty}\!\!{dp_z\over2\pi}{1\over{2\varepsilon_n}}e^{-i\left[\varepsilon_n-\Omega \left(l+{1\over2}\right)\right](t-t^\prime)+ip_z(z-z^\prime)}
H_{n,l}(\theta,r)\left[\sum_{{\rm s}=\pm}u_{n,l}^s(p_z)u_{n,l}^{s\dagger}(p_z)\right]H_{n,l}^\dagger(\theta',r')\gamma^0\nonumber\\
&=&\!\sum_{n=0}^\infty\!\sum_l\!\!\!\int_{-\infty}^{\infty}\!\!{dp_z\over2\pi}{e^{-i\left[\varepsilon_n-\Omega \left(l+{1\over2}\right)\right](t-t^\prime)+ip_z(z-z^\prime)}\over{2\varepsilon_n}}
\Bigg\{\left[{\cal P}_\uparrow\chi_{n,l}(\theta,r)\chi_{n,l}^*(\theta',r')\!+\!{\cal P}_\downarrow\chi_{n-1,l+1}(\theta,r)\chi_{n-1,l+1}^*(\theta',r')\right]\nonumber\\
&&(\gamma^0\varepsilon_n-\gamma^3p_z+m)-\left[{\cal P}_\uparrow\chi_{n,l}(\theta,r)\chi_{n-1,l+1}^*(\theta',r')+{\cal P}_\downarrow\chi_{n-1,l+1}(\theta,r)\chi_{n,l}^*(\theta',r')\right]\sqrt{2n qB}\gamma^2\Bigg\}.
\end{eqnarray}
Accordingly, the Feynman Green's function can be given by
\begin{eqnarray}\label{propogator}
	S_{\rm F}(x,x')&\equiv&\langle0|{\rm T}\psi(x)\bar{\psi}(x^\prime)|0\rangle\nonumber\\
	&=&\sum_{n=0}^\infty\sum_l\int\int{dp_0dp_z\over(2\pi)^2}{i~e^{-ip_0(t-t^\prime)+ip_z(z-z^\prime)}\over\left({p}_0^{l+}\right)^2-\varepsilon_n^2+i\epsilon}
	\Bigg\{\left[{\cal P}_\uparrow\chi_{n,l}(\theta,r)\chi_{n,l}^*(\theta',r')+{\cal P}_\downarrow\chi_{n-1,l+1}(\theta,r)\chi_{n-1,l+1}^*(\theta',r')\right]\nonumber\\
	&&\left(\gamma^0{p}_0^{l+}-\gamma^3p_z+m\right)-\left[{\cal P}_\uparrow\chi_{n,l}(\theta,r)\chi_{n-1,l+1}^*(\theta',r')+{\cal P}_\downarrow\chi_{n-1,l+1}(\theta,r)\chi_{n,l}^*(\theta',r')\right]\sqrt{2n qB}\gamma^2\Bigg\}
\end{eqnarray}
\end{widetext}	
with ${p}_0^{ls}=p_0+\Omega \left(l+s{1\over2}\right)$. It is easy to show that this propagator can be derived straightforwardly from the one in the absence of rotation ($\Omega=0$) by a coordinate shift $\theta\rightarrow\theta+\Omega t$ and including the spin-rotation coupling ${1\over2}\Omega$.  The lowest Landau level (LLL) contribution comes from the term $\chi_{n,l}(\theta,r)\chi_{n,l}^*(\theta',r')$ as should be.

 The correctness of the above fermion propagator can be briefly checked in the vanishing rotation limit, where only the function $\chi_{n,l}(\theta,r)$ and its conjugate depend on the orbital angular quantum number $l$. For the lowest Landau level with $n=0$, the summation over $l$ can be completed as
\begin{eqnarray}\label{L0}
&&\sum_{l=0}^\infty\chi_{0,l}(\theta,r)\chi_{0,l}^*(\theta',r')\nonumber\\
&=&\sum_{l=0}^\infty{{qB}\over l!}{e^{i\,l(\theta-\theta')}\over2\pi}\left({qB\over2}rr'\right)^le^{-{qB\over4}(r^2+{r'}^2)}\nonumber\\
&=&{qB\over2\pi}e^{iq\int_x^{x'} A_\mu dx_\mu-{qB\over4}(\mathbf{r}-\mathbf{r'})^2}
\end{eqnarray}
with the exponent in Schwinger phase $iq\int_x^{x'} A_\mu dx_\mu=i{qB\over2}\sin(\theta-\theta')rr'$. For the first Landau level with $n=1$, we have
\begin{eqnarray}\label{L1}
&&\sum_{l=-1}^\infty\chi_{1,l}(\theta,r)\chi_{1,l}^*(\theta',r')\nonumber\\
&=&\sum_{l=-1}^\infty{{qB}\over (l+1)!}{\left(e^{i\,(\theta-\theta')}{qB\over2}rr'\right)^l\over2\pi ~e^{{qB\over4}(r^2+{r'}^2)}}L_1^{l}\left(\tilde{r}^2\right)L_1^{l}\left({\tilde{r}^{\prime2}}\right)\nonumber\\
&=&{qB\over2\pi}e^{iq\int_x^{x'} A_\mu dx_\mu-{qB\over4}(\mathbf{r}-\mathbf{r'})^2}\left[1-{qB\over2}(\mathbf{r}-\mathbf{r'})^2\right].
\end{eqnarray}
We find that the summations over $l$ give the well-known Schwinger phase consistently.  Apart from the Schwinger phase, by taking Fourier transformation of Eqs.(\ref{L0}) and (\ref{L1}), we can exactly reproduce the corresponding terms of the effective fermion propagator in Euclidean space~\cite{Miransky:2015ava}.  However, in the presence of rotation,  the situation becomes quite different. Because $p_0^{l+}$ depends on $l$, the summation over $l$ does not give rise to a simple Schwinger phase. Therefore, the full propagator cannot be decomposed into a translation invariant part multiplying by a general Schwinger phase. 

For the case $qB<0$, the solution can be obtained by making the replacements $H_{n,l}\rightarrow H_{n,l}^{-}$ and $\chi_{n,l}\rightarrow\chi_{n,l}^{-}$, where
\begin{eqnarray}
&&H_{n,l}^{-}(\theta,r)=\left[{\cal P}_\uparrow\chi_{n-1,l-1}^{-}(\theta,r)+{\cal P}_\downarrow\chi_{n,l}^{-}(\theta,r)\right],\nonumber\\
&&\chi_{n,l}^{-}(\theta,r)=\left({|qB|\over2\pi}{ n!\over(n-l)!}\right)^{1\over2}{e^{i\, l\theta}}~\tilde{r}^{-l}e^{-\tilde{r}^2/2}L_n^{-l}\left(\tilde{r}^2\right).\nonumber
\end{eqnarray}
Here the Laguerre polynomial $L_n^{-l}(x)$ is nonvanishing only for $n\ge0$ and $-l\in\left[-n,N-n\right]$, with $N=\left \lfloor|qB|S/(2\pi) \right \rfloor$.
The eigenvalue equation for $u_{n,l}(p_z)$ reads
\begin{eqnarray}
\left(\gamma^0\varepsilon^--\gamma^3p_z+\gamma^2\sqrt{2n |qB|}-m\right)u_{n,l}(p_z)=0
\end{eqnarray}
with $\varepsilon^-\equiv E+\Omega \left(l-{1\over2}\right)$, which can be solved in a similar way as the case $qB>0$. The Feynman Green's function is finally given by
\begin{widetext}
	\begin{eqnarray}\label{qn}
	S_{\rm F}(x,x')&=&\sum_{n=0}^\infty\sum_l\int_{-\infty}^{\infty}{dp_0dp_z\over(2\pi)^2}{i~e^{-ip_0(t-t^\prime)+ip_z(z-z^\prime)}\over\left({p}_0^{l-}\right)^2-\varepsilon_n^2+i\epsilon}
	\left\{\left[{\cal P}_\uparrow\chi_{n-1,l-1}^{-}(\theta,r)\chi_{n-1,l-1}^{-*}(\theta',r')+{\cal P}_\downarrow\chi_{n,l}^{-}(\theta,r)\chi_{n,l}^{-*}(\theta',r')\right]\right.\nonumber\\
	&&\left.\left(\gamma^0{p}_0^{l-}-\gamma^3p_z+m\right)+\left[{\cal P}_\uparrow\chi_{n-1,l-1}^{-}(\theta,r)\chi_{n,l}^{-*}(\theta',r')+{\cal P}_\downarrow\chi_{n,l}^{-}(\theta,r)\chi_{n-1,l-1}^{-*}(\theta',r')\right]\sqrt{2n |qB|}\gamma^2\right\}.
	\end{eqnarray}	
\end{widetext}

\subsection{Gap equation for dynamical quark mass}\label{gapcoefficient}

Now we investigate the dynamical quark mass induced by the four fermion interaction under the circumstance of PRM.  Here we first consider the state with vanishing pion condensate, i.e.,
\begin{eqnarray}\label{VPC}
\langle\bar{\psi}\psi\rangle\neq0, \ \ \  \langle\bar{\psi}i\gamma_5\mbox{\boldmath{$\tau$}}\psi\rangle=0.
\end{eqnarray}
In the mean-field approximation, the effective action is given by
\begin{eqnarray}\label{Actionm}
\Gamma_{\rm eff}(m)&=&\int d^4x\frac{(m-m_0)^2}{4G}\nonumber\\
&&-\ i\ln{\rm Det}\left[i\gamma^\mu(D_\mu\!+\!\Gamma_\mu)\!-\!m\right],
\end{eqnarray}
where dynamical quark mass $m$ is defined as
 \begin{eqnarray}
m=m_0-2G\langle\bar{\psi}\psi\rangle.
\end{eqnarray}
In the presence of rotation, the area $S$ of the $xy$-plane cannot be infinitely large.  We set $S=\pi R^2$  with $R$ being the radius of the $xy$-plane, then causality requires that $\Omega R<1$. Due to the
existence of a boundary $(r=R)$, the chiral condensate or the dynamical quark mass becomes generically inhomogeneous, i.e., $m=m(r)$. From the variational condition, $\delta \Gamma_{\rm eff}/\delta m=0$, we obtain the gap equation~\cite{Wang:2019nhd}
 \begin{eqnarray}
m(r)-m_0=2G{\rm Tr} \left[S_{\rm F}(x,x)\right].
\end{eqnarray}
Generalization to finite temperature is straightforward, by using the standard imaginary time formalism.

Considering the spatial dependence of the dynamical quark mass is rather complicated.  For sufficiently large $R$, the inhomogeneity appears only in a narrow regime near the boundary $r=R$~\cite{Chen:2015hfc,Wang:2019nhd,NJL-Rotation01,NJL-Rotation02,NJL-Rotation03}.  As a good approximation, we treat $m$ as a constant. The effective action and the gap equation can be conveniently evaluated. We employ a regularization scheme where the divergent vacuum part is explicitly separated out~\cite{Cao:2015xja,Cao:2014uva}. The gap equation at finite temperature is explicitly given by
\begin{widetext}
\begin{eqnarray}\label{gap}
{m-m_0\over 2GN_c}&=&{m^2\over\pi^2}\left[\Lambda\sqrt{1+{\Lambda^2\over m^2}}-m\ln\left({\Lambda\over m}
+\sqrt{1+{\Lambda^2\over m^2}}\right)\right]+\frac{m}{4\pi^2}\sum_{{\rm f}=u,d}\int_0^\infty{{\rm d}s\over s^{2}} e^{-sm^2}\left({q_{\rm f}Bs\over\tanh{q_{\rm f}Bs}}-1\right)\nonumber\\
&&-\ m\sum_{{\rm f}=u,d}\sum_{n=0}^\infty{1\over S}\sum_{l=0}^{N_{{\rm f}}}\int_{-\infty}^{\infty}{dp_z\over\pi}{\alpha_n\over\varepsilon_{n{\rm f}}}
\Big[f(\varepsilon_{n{\rm f}}+\Omega_{nl})+f(\varepsilon_{n{\rm f}}-\Omega_{nl})\Big],
\label{masseqn1}
	\end{eqnarray}
\end{widetext}
where $N_c=3$ is the color degree of freedom, $\Lambda$ is the three-momentum cutoff as in the vacuum case, $\alpha_n=(2-\delta_{n0})/2$, $\Omega_{nl}=\left(l-n+1/2\right)\Omega$, the cutoff for the $l$-sum is given by $N_{{\rm f}}=\left \lfloor|q_{\rm f}B|S/(2\pi) \right \rfloor$, and $f(E)=1/(e^{ E/T}+1)$ is the Fermi-Dirac distribution function with $T$ the temperature. For vanishing rotation $\Omega=0$, the $l$-sum ${1\over S}\sum_l$ in the last term just gives the well-known degeneracy factor $|q_{\rm f}B|/(2\pi)$. However, for nonzero rotation $\Omega\neq0$, states with different $l$ are no longer degenerate due to the quantity $\Omega_{nl}$ in the Fermi-Dirac distribution function, which acts like a baryon chemical potential. The $l$-sum is divergent if we take $R\rightarrow\infty$. Here we consider a finite-size cylindrical system, of which the radius satisfies $1/\sqrt{|q_{\rm f}B|}\ll R\leq 1/\Omega$~\cite{Chen:2015hfc}. 

For numerical calculations, we use the model parameter set: $m_0=5~$MeV, $\Lambda=653~$MeV, $G=4.93~$GeV$^{-2}$ and study the cases where the magnetic field is sufficiently strong: $eB=0.5~$GeV$^2$ and $eB=1~$GeV$^2$ for illuminations. In Fig.~\ref{B05}(a) and Fig.~\ref{B1}(a), we show the dependence of the dynamical mass $m$ on the angular velocity $\Omega$. In the calculations, we considered two values of the system size: $R_1=20/\sqrt{2eB}$ and $R_2=20/\sqrt{eB}$, which are of a few fermi.  The results are qualitatively consistent with the findings in \cite{Chen:2015hfc}: The dynamical quark mass keeps a constant for small angular velocity and starts to decrease at a critical angular velocity $\Omega=\Omega_m$. For the magnetic field and system size we choose, $\Omega_m$ is of order $O({\rm MeV})$.

\section{Stability against charged pion condensation}\label{numerical}
In this section, we explore the possibility of charged pion condensation induced by rotation in a strong magnetic field. To this end, we consider the state with a nonzero dynamical mass and charged pion condensate, i.e.,
\begin{eqnarray}
\langle\bar{\psi}\psi\rangle\neq0, \ \ \  \langle\bar{\psi}i\gamma_5\tau_\pm\psi\rangle\neq0,
\end{eqnarray}
where $\tau_\pm=\tau_1\pm i\tau_2$. For convenience, we define the CPC order parameter as
\begin{eqnarray}
\Delta=2G \langle\bar{\psi}i\gamma_5\tau_+\psi\rangle=4G\langle\bar{u}i\gamma_5d\rangle.
\end{eqnarray}
Because of the residual isospin U$(1)$ symmetry, $\Delta$ can be chosen to be real without loss of generality.

\begin{figure}[!htb]
	\centering
	\includegraphics[width=0.42\textwidth]{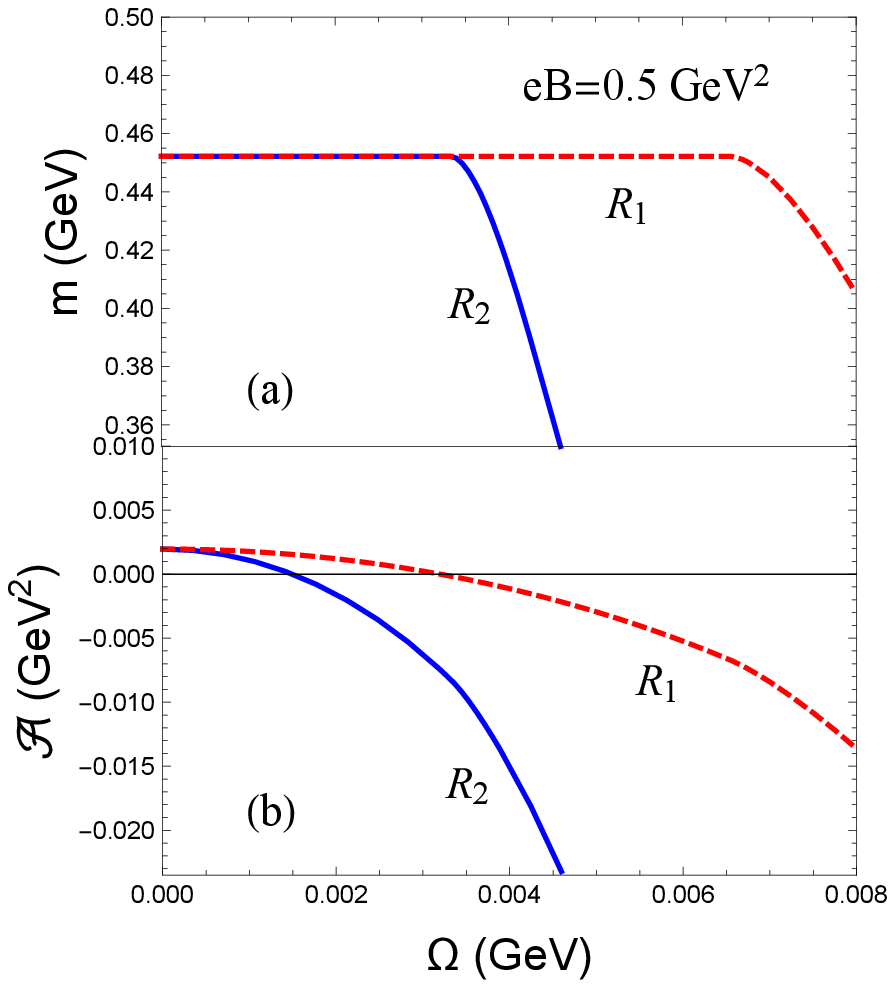}
	\caption{The dependence of the dynamical quark mass $m$ (a) and the coefficient ${\cal A}$ (b) on the angular velocity $\Omega$ for two system sizes $R_1=20/\sqrt{2eB}$ (red dashed lines) and $R_2=20/\sqrt{eB}$ (blue solid lines) at magnetic field $eB=0.5~{\rm GeV}^2$.}\label{B05}
\end{figure}

\begin{figure}[!htb]
	\centering
	\includegraphics[width=0.425\textwidth]{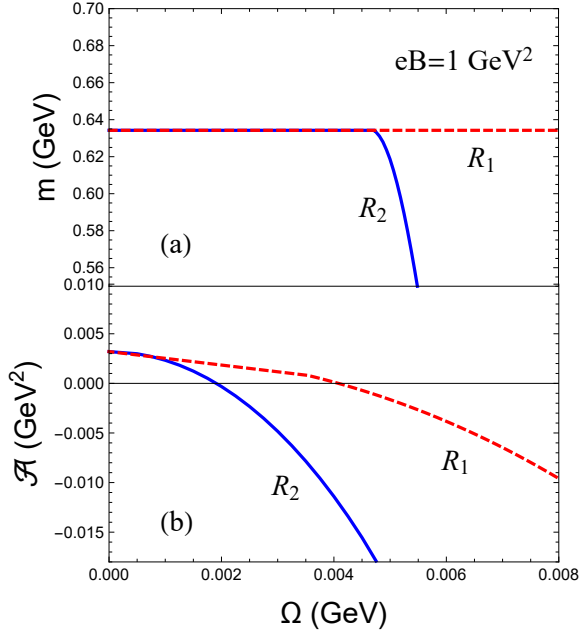}
	\caption{The dependence of the dynamical quark mass $m$ (a) and the coefficient ${\cal A}$ (b) on the angular velocity $\Omega$ for two system sizes $R_1=20/\sqrt{2eB}$ (red dashed lines) and $R_2=20/\sqrt{eB}$ (blue solid lines) at magnetic field $eB=1~{\rm GeV}^2$.}\label{B1}
\end{figure}

The brute-force method is to derive coupled gap equations for $m$ and $\Delta$, and solve them simultaneously.
However, because the charged pion condensate couples different Landau levels of $u$ and $d$ quarks, the corresponding quark propagator matrix mixes all Landau levels and the calculation becomes rather complicated~\cite{Cao:2015xja}. Thus we study the stability of the system against formation of CPC based on the state given by Eqs.(\ref{VPC}) and (\ref{gap}). The effective potential can be expressed as the form in Ginzburg-Landau theory
\begin{eqnarray}\label{thermo}
V_{\rm eff}(m,\Delta)=V_{\rm eff}(m,0)+{\cal A}~\Delta^2+{\cal B}~\Delta^4+\dots,
\end{eqnarray}
where $V_{\rm eff}(m,0)$ is given by Eq.(\ref{Actionm}) divided by the volume of the system. In a pure magnetic field, it has been shown that the coefficient ${\cal B}$ of the quartic term is positive ~\cite{Cao:2015xja}. As we will see in the numerical results, the transition to the CPC occurs at relatively small angular velocity $\Omega$, so we can set ${\cal B}>0$ in our study. The coefficient  ${\cal A}$ in the quadratic term thus characterizes the stability against formation of CPC. If ${\cal A}<0$, the true ground state of the system prefers a nonzero charged pion condensate.

In the absence of rotation, the coefficient ${\cal A}$ is given by 
$${\cal A}={1\over4G}+{\cal A}_{FL}$$
with the contribution from fermion loop as
\begin{equation}\label{AFL0}
{\cal A}_{FL}={i\over V_4}{\rm Tr}\left[S_{\rm F}^u(x,y)i\gamma^5S_{\rm F}^d(y,x)i\gamma^5e^{-ie\int_x^y{A}_\mu(z) d{z}_\mu}\right],
\end{equation}
where $V_4$ is the space-time volume and the trace is taken over the internal and coordinate spaces.
Note that we have compensated the gauge link contribution (the phase term) from the charged pion condensate in order to keep ${\cal A}$ gauge invariant~\cite{Cao:2015xja}. This process is equivalent to define a local condensate with a Wilson line: 
\begin{eqnarray}
\tilde{\Delta}(y)=e^{-ie\int_0^y[{A}_\mu(z)+{1\over2}F_{\mu\nu}{z}^\nu] d{z}_\mu}\Delta,
\end{eqnarray}
which guarantees that the effective potential $V_{\rm eff}(m,\Delta)$ is formally gauge invariant and we can expand it in powers of the gauge-independent quantity $\Delta$. When rotation is turned on, the gauge link contribution may be ambiguous because we cannot read out a general phase term that does not only depend on the relative distance between $u$ and $d$ quarks in the Minkowski coordinate system from the pion propagator. Taking LLL for example, the gauge-link term in pion propagator looks like that in Eq.(\ref{L0}) but with the polar angle shifted: $\theta\rightarrow\theta+\Omega t$. The same argument also applies to higher Landau levels. Nevertheless, the integral path in the gauge link can be shown to be geodesic; see Ref.~\cite{Chen:2019tcp} for more detailed discussions. 

In this work, we consider the case that the magnetic field is sufficiently strong ($\sqrt{eB}\sim1{\rm GeV}$) and the rotation is relatively slow ($\Omega$ of order a few MeV). Thus, we can assume that the charged pions condense in the static Landau state in the laboratory frame and hence does not feel the rotation, which is in analogy to that found in $^3$He superfluid for small rotation~\cite{he3}.  In this case, the coefficient ${\cal A}_{FL}$ is also given by Eq.(\ref{AFL0}), but with the induced scalar potential $A_0=-B\Omega r^2/2$ neglected, i.e., 
$A_\mu=(0,By/2,-Bx/2,0)$ as in the laboratory frame.  For this static Landau state of the charged pion condensate, the coefficient ${\cal A}_{FL}$ can be evaluated as 
\begin{widetext}
\begin{eqnarray}\label{AFL}
{\cal A}_{FL}
&=&-{N_c\over 2S}\sum_{n=0}^\infty\sum_{l=0}\sum_{n'=0}^\infty\sum_{l'=0}\sum_{s=\pm}\int_{-\infty}^{\infty}{dp_z\over(2\pi)}{\tanh\left({\varepsilon_n^u-s~\Omega(l-n+{1\over2})\over2T}\right){n!n'!\over l!l'!}\left({q_uB\over 2}\right)^{l-n+1}\left({|q_dB|\over 2}\right)^{l'-n'+1}\over\varepsilon_n^u\left[\left(\varepsilon_n^u-s~\Omega_{nl,n'l'}\right)^2-(\varepsilon_{n'}^d)^2\right]}\nonumber\\
&&\Bigg\{\Bigg[{|q_dB|\over n'}G_{nl,n'l'}(q_uB,|q_dB|)+{q_uB\over n}G_{n'l',nl}(|q_dB|,q_uB)\Bigg]\left[\left(\varepsilon_n^u-s~\Omega_{nl,n'l'}\right)^2-(\varepsilon_{n'}^d)^2-\Omega_{nl,n'l'}^2\right]\nonumber\\
&&+H_{nl,n'l'}(q_uB,|q_dB|)\Bigg\}+\left(\varepsilon_{n'}^d\leftrightarrow\varepsilon_{n}^u,nl\leftrightarrow n'l',q_u\leftrightarrow |q_d|\right),
\end{eqnarray}
where  $\varepsilon_n^u=(p_z^2+2nq_uB+m^2)^{1/2}$, $\varepsilon_n^d=(p_z^2+2n|q_dB|+m^2)^{1/2}$, and $\Omega_{nl,n'l'}=(l+l'-n-n'+1)\Omega$. The details of the calculations and the definitions of the auxiliary functions $G_{nl,n'l'}(q_uB,|q_dB|)$ and $H_{nl,n'l'}(q_uB,|q_dB|)$ can be found in Appendix~\ref{coefficient}.
Again, following the ``vacuum regularization" scheme~\cite{Cao:2014uva}, the coefficient ${\cal A}$ can be divided into three parts: ${\cal A}={\cal A}_{0}+{\cal A}_{\rm B}+{\cal A}_\Omega$, where ${\cal A}_{0}$ and ${\cal A}_{\rm B}$ are the contributions from the vacuum and the pure magnetic field effect, respectively~\cite{Cao:2015xja}. We have
\begin{eqnarray}\label{A0B}
{\cal A}_0 &=&{1\over4G}-{N_c\Lambda^2\over 2\pi^2}\Bigg[\sqrt{1+{m^2\over \Lambda^2}}-{m^2\over \Lambda^2}\ln\left({\Lambda\over m}
+\sqrt{1+{\Lambda^2\over m^2}}\right)\Bigg],\\
{\cal A}_{B}&=&-{N_c\over 8\pi^2}\int_0^\infty {e^{-s\,m^2}ds\over s^2}\int_{-1}^1 dv\left\{\left[{(1\!-\!g_{\rm u}^2(s,v))(1\!-\!g_{\rm d}^2(s,v))\over \left({g_{\rm u}(s,v)\over q_{\rm u}Bs}+{g_{\rm d}(s,v)\over q_{\rm d}Bs}\right)^2}-1\right]+\left(s m^2\!+\!1\right)\left[{1\!+\!g_{\rm u}(s,v)g_{\rm d}(s,v)\over {g_{\rm u}(s,v)\over q_{\rm u}Bs}+{g_{\rm d}(s,v)\over q_{\rm d}Bs}}-1\right]\right\},
\end{eqnarray}
where $g_{\rm u}(s,v)\equiv\tanh ({1+ v\over2}q_{\rm u}Bs)$ and $g_{\rm d}(s,v)\equiv\tanh ({1-v\over2}q_{\rm d}Bs)$. The rotation contribution as well as the temperature effect is included in ${\cal A}_\Omega$ which is explicitly
\begin{eqnarray}\label{AO}
{\cal A}_\Omega&=&-{N_c\over 2S}\sum_{n=0}^\infty\sum_{l=0}^{N_u}\sum_{n'=0}^\infty\sum_{l'=0}^{N_d}{{n!n'!\over l!l'!}\!\!\left({q_uB\over 2}\right)^{l-n+1}\!\!\!\left({|q_dB|\over 2}\right)^{l'-n'+1}}\left\{\Bigg[{|q_dB|\over n'}G_{nl,n'l'}(q_uB,|q_dB|)\!+\!{q_uB\over n}G_{n'l',nl}(|q_dB|,q_uB)\Bigg]\right.\nonumber\\
&&\sum_{s=\pm}\int_{-\infty}^{\infty}{dp_z\over(2\pi)}\left[ {\tanh\left({\varepsilon_n^u-s~\Omega(l-n+{1\over2})\over2T}\right)-1\over\varepsilon_n^u}-{\Omega_{nl,n'l'}^2\tanh\left({\varepsilon_n^u-s~\Omega(l-n+{1\over2})\over2T}\right)\over\varepsilon_n^u\left[\left(\varepsilon_n^u-s~\Omega_{nl,n'l'}\right)^2-(\varepsilon_{n'}^d)^2\right]}\right]+H_{nl,n'l'}(q_uB,|q_dB|)\nonumber\\
&&\sum_{s=\pm}\left.\int_{-\infty}^{\infty}{dp_z\over(2\pi)}{1\over\varepsilon_n^u }\left[{\tanh\left({\varepsilon_n^u-s~\Omega(l-n+{1\over2})\over2T}\right)\over\left(\varepsilon_n^u-s~\Omega_{nl,n'l'}\right)^2-(\varepsilon_{n'}^d)^2}-{1\over\left(\varepsilon_n^u\right)^2-(\varepsilon_{n'}^d)^2}\right]\right\}+\left(\varepsilon_{n'}^d\leftrightarrow\varepsilon_{n}^u,nl\leftrightarrow n'l',q_u\leftrightarrow |q_d|\right).
\end{eqnarray}
\end{widetext}

The coefficient ${\cal A}$ can be computed once we have solved the dynamical mass $m$ from the gap equation Eq.(\ref{gap}). Since the magnetic field is strong, we can analytically work out the functions
 $G_{nl,n'l'}(q_uB,|q_dB|)$ and $H_{nl,n'l'}(q_uB,|q_dB|)$ up to the $10$-th Landau levels ($n_{\rm max}=n'_{\rm max}=10$) by using {\it Mathematica}. We have checked that the truncations of the Landau level summations give very accurate value of the coefficient ${\cal A}$ for the chosen magnetic fields $eB=0.5~{\rm GeV}^2$ and $eB=1~{\rm GeV}^2$.  Again, we also consider two system sizes $R_1=20/\sqrt{2eB}$ and $R_2=20/\sqrt{eB}$.
 
The dependence of the coefficient ${\cal A}$ on the angular velocity $\Omega$ is shown in Fig.~\ref{B05}(b) and Fig.~\ref{B1}(b).  For the magnetic fields and system sizes we considered, we find that the coefficient ${\cal A}$ becomes negative if $\Omega$ exceeds a critical value $\Omega_{\rm PC}$, which is of order $O(1~{\rm MeV})$.  In the regime with ${\cal A}<0$, CPC becomes energetically favored. Another important observation is that 
the critical angular velocity $\Omega_{\rm PC}$ at which CPC occurs is much smaller than another critical value $\Omega_m$ observed in the last section at which the dynamical mass $m$ starts to decrease if only the chiral condensate is considered. As a result, if the CPC is taken into account, the chiral condensate or the dynamical mass $m$ will start to decrease at $\Omega_{\rm PC}$. This phenomenon is quite similar to the case with finite isospin chemical potential $\mu_I$, where the CPC occurs and the dynamical quark mass $m$ gets reduced at the critical isospin chemical potential equal to the vacuum pion mass~\cite{He2005}. Without considering the CPC, $m$ will start to decrease at $\mu_I=2m$ according to the Silver-Blaze theorem~\cite{Cohen:2003kd}, just like the case with baryon chemical potential. It is also found that for a given value of the magnetic field, a larger system size leads to a smaller value of $\Omega_{\rm PC}$. The reason is that the quantity $\Omega_{nl,n'l'}$ plays the role of an effective isospin chemical potential and the upper limits for the $l$- and $l'$-sums become also larger for larger system size.

The final question is that how we can reconcile the contradiction between the initial expectation from spin argument which disfavors CPC and the above numerical results which shows that the CPC becomes energetically favored beyond $\Omega_{\rm PC}$. The key point is that our initial spin argument is based on the LLL approximation where only one option of the spin is available for each quark flavor. Actually, for LLL, we have $G_{0l,0l'}=H_{0l,0l'}=0$ regardless the values of $l$ and $l'$, then ${\cal A}_\Omega=0$ in the LLL approximation and ${\cal A}$ is positive definite. In this case we may conclude that the CPC is not favored, consistent with our initial expectation. However, the contribution from higher Landau levels, with both spin up and down components for each quark, are also important. From the explicit form of ${\cal A}_\Omega$ in Eq.(\ref{AO}), we find that the rotation induces the effect of isospin chemical potential, i.e., the quantity $\Omega_{nl,n'l'}$ plays the role of an effective isospin chemical potential~\cite{He2005}. Therefore, there exist two competing effects: the spin breaking effect functioning through the LLL and the isospin enhancement effect functioning through the higher Landau levels. The numerical calculations indicate that the later overwhelms the former, leading to CPC when a sufficiently rapid rotation is turned on.

\section{Summary}\label{conclusions}
Based on the non-interacting Klein-Gordon theory of charged pions, the previous study of QCD system predicted that charged pions would get condensed for sufficiently rapid rotation in a strong parallel magnetic field, which can be realized in relativistic heavy ion collisions~\cite{Liu:2017spl}. However, it is known that a strong magnetic field will significantly influence the meson properties through the internal quark structures of the mesons. Actually, exploring the internal quark structures of the charged pions, the spin breaking effect may disfavor the condensation of charged pions. Therefore, it is important to check this prediction for charged pion condensation by adopting an interacting theory with quarks as elementary degrees of freedom. In this work, we have studied the stability of the QCD system under PRM against the formation of CPC within the Nambu--Jona-Lasino model. Technically, we adopted a Ginzburg-Landau-like approach and evaluated the coefficient of the quadratic term which characterizes the stability against CPC.  Quantitatively, we observed the spin breaking effect functioning through the LLL and the isospin enhancement effect functioning through the higher Landau levels. The later overwhelms the former and hence CPC becomes energetically favored when a sufficiently rapid rotation is turned on.

It is significant to mention that the values of the magnetic field and the angular velocity we used in this work are all reachable in peripheral heavy ion collisions. Actually, the chosen magnetic field is almost the strongest that can be produced in heavy ion experiments and the fastest rotation was found to be $\Omega\approx(9\pm1)\times10^{21}~{\rm Hz}\sim6~{\rm MeV}$~\cite{STAR:2017ckg}. Therefore, we wish that the CPC will be explored in the future experiments. On the theoretical side, the structure of the charged pion condensate and the corresponding critical temperature still need further investigations. With increasing rotation velocity, the charged pion condensate could not be in a uniform state but rather forms some interesting vortex lattice structure as in the $^3$He system~\cite{he3}. We expect that the Bogoliubov-de Gennes method developed for rotating finite-size system~\cite{Wang:2019nhd} can be applied to study the CPC.

\emph{Acknowledgments}---
We thank Haolei Chen, Xu-Guang Huang and Kazuya Mameda for useful discussions and communications. G.C. is supported by the National Natural Science Foundation of China with Grant No. 11805290 and Young Teachers Training Program of Sun Yat-sen University with Grant No. 19lgpy282. L. H. acknowledges the support from the National Natural Science Foundation of China, Grant Nos. 11775123 and 11890712.

\appendix

\begin{widetext}
	\section{Calculation of the coefficient ${\cal A}_{FL}$}\label{coefficient}
	Using the fermion propagator derived in Sec. II, the coefficient ${\cal A}_{FL}$ can be expressed as
	
	\begin{eqnarray}
	{\cal A}_{FL}&=&{-i\over S}\sum_{n=0}^\infty\sum_l\sum_{n'=0}^\infty\sum_{l'}\int_{-\infty}^{\infty}{dp_0\over2\pi}\int_{-\infty}^{\infty}{dp_z\over2\pi}
	{\rm Tr}\Bigg\{\left[{\cal P}_\uparrow\chi_{n,l}(\theta,r)\chi_{n,l}^*(\theta',r')+{\cal P}_\downarrow\chi_{n-1,l+1}(\theta,r)\chi_{n-1,l+1}^*(\theta',r')\right]\nonumber\\
	&&\ \ \ \ \times\left(\gamma^0{p}_0^{l+}-\gamma^3p_z+m\right)-\left[{\cal P}_\uparrow\chi_{n,l}(\theta,r)\chi_{n-1,l+1}^*(\theta',r')+{\cal P}_\uparrow\chi_{n-1,l+1}(\theta,r)\chi_{n,l}^*(\theta',r')\right]
	\gamma^2\sqrt{2n q_uB}\Bigg\}\nonumber\\
	&&\times\Bigg\{\left[{\cal P}_\uparrow\chi_{n'-1,l'-1}^{-}(\theta',r')\chi_{n'-1,l'-1}^{-*}(\theta,r)+{\cal P}_\downarrow\chi_{n',l'}^{-}(\theta',r')\chi_{n',l'}^{-*}(\theta,r)\right]\left(\gamma^0{p}_0^{l'-}-\gamma^3p_z-m\right)\nonumber\\
	&&\ \ \ \ \  +\left[{\cal P}_\uparrow\chi_{n'-1,l'-1}^{-}(\theta',r')\chi_{n',l'}^{-*}(\theta,r)+{\cal P}_\downarrow\chi_{n',l'}^{-}(\theta',r')\chi_{n'-1,l'-1}^{-*}(\theta,r)\right]\gamma^2\sqrt{2n'|q_dB|}\Bigg\}\nonumber\\
	&&\times{e^{-ie\int_x^y{A}_\mu(z) d{z}_\mu}\over\left[\left({p}_0^{l+}\right)^2-(\varepsilon_n^u)^2\right]\left[\left({p}_0^{l'-}\right)^2-(\varepsilon_{n'}^d)^2\right]},
	\end{eqnarray}
	where the trace is taken over all internal spaces and the coordinate space. Completing the traces in the internal spaces, we obtain
	\begin{eqnarray}
	{\cal A}_{FL}&=&{-2N_ci\over S}\sum_{n=0}^\infty\sum_l\sum_{n'=0}^\infty\sum_{l'}\sum_{r,r'}\sum_{\theta,\theta'}\int_{-\infty}^{\infty}{dp_0\over2\pi}\int_{-\infty}^{\infty}{dp_z\over2\pi}
	{e^{-ie\int{A}_\mu(z) dz_\mu}\over\left[\left({p}_0^{l+}\right)^2-(\varepsilon_n^u)^2\right]\left[\left({p}_0^{l'-}\right)^2-(\varepsilon_{n'}^d)^2\right]}\nonumber\\
	&&\Bigg\{\left[\chi_{n,l}(\theta,r)\chi_{n,l}^*(\theta',r')\chi_{n'-1,l'-1}^{-}(\theta',r')\chi_{n'-1,l'-1}^{-*}(\theta,r)+\chi_{n-1,l+1}(\theta,r)\chi_{n-1,l+1}^*(\theta',r')\chi_{n',l'}^{-}(\theta',r')\chi_{n',l'}^{-*}(\theta,r)\right]\nonumber\\
	&&\ \ \ \times\left({p}_0^{l+}{p}_0^{l'-}-p_z^2-m^2\right)+2\sqrt{(2nq_uB)2n'|q_dB|}\chi_{n,l}(\theta,r)\chi_{n-1,l+1}^*(\theta',r')\chi_{n',l'}^{-}(\theta',r')\chi_{n'-1,l'-1}^{-*}(\theta,r)\Bigg\},
	\end{eqnarray}
	where $\sum_{r,r'}=\int_0^\infty rdr\int_0^\infty r'dr'$ and $\sum_{\theta,\theta'}=\int_0^{2\pi}d\theta\int_0^{2\pi}d\theta'$. Then, the integrals over the polar angles can be completed to give
	\begin{eqnarray}
	{\cal A}_{FL}&=&{-4N_ci\over S}\sum_{n=0}^\infty\sum_{l=0}\sum_{n'=0}^\infty\sum_{l'=0}\sum_{r,r'}\int_{-\infty}^{\infty}{dp_0\over2\pi}\int_{-\infty}^{\infty}{dp_z\over2\pi}
	{J_{l+l'-n-n'+1}\left({eB\over2}rr'\right)(rr')^{l+l'-n-n'+1}e^{-eB(r^2+{r'}^2)/4}\over\left[\left({p}_0^{(l-n)+}\right)^2-(\varepsilon_n^u)^2\right]\left[\left({p}_0^{(n'-l')-}\right)^2-(\varepsilon_{n'}^d)^2\right]}\nonumber\\
	&&{n!n'!\over l!l'!}\left({q_uB\over 2}\right)^{l-n+1}\left({|q_dB|\over 2}\right)^{l'-n'+1}\Bigg\{\bigg[{|q_dB|\over n'}{F}_{nl,n'l'}(q_uB,|q_dB|;r){F}_{nl,n'l'}(q_uB,|q_dB|;r')\nonumber\\
	&&+{q_uB\over n}{F}_{n'l',nl}(|q_dB|,q_uB;r){F}_{n'l',nl}(|q_dB|,q_uB;r')\bigg]\left({p}_0^{(l-n)+}{p}_0^{(n'-l')-}-p_z^2-m^2\right)\nonumber\\
	&&+4{|q_dB|}{q_uB}{F}_{nl,n'l'}(q_uB,|q_dB|;r){F}_{n'l',nl}(|q_dB|,q_uB;r')\Bigg\},
	\end{eqnarray}
	where the function $F$ is defined as
	\begin{eqnarray}
	F_{nl,n'l'}(q_uB,|q_dB|;x)\equiv L_n^{l-n}\left({q_uB ~x^2\over2}\right)L_{n'-1}^{l'-n'+1}\left({|q_dB| ~x^2\over2}\right).
	\end{eqnarray}	
	Using this function and further defining auxiliary functions: 
	\begin{eqnarray}
	G_{nl,n'l'}(q_uB,|q_dB|)&\equiv&\int_0^\infty{\di r \di r'}~J_{l+l'-n-n'+1}\left({eB\over2}rr'\right)(rr')^{l+l'-n-n'+2}e^{-eB(r^2+{r'}^2)/4}\prod_{x=r,r'}{F}_{nl,n'l'}(q_uB,|q_dB|;x)\nonumber\\
	H_{nl,n'l'}(q_uB,|q_dB|)&\equiv&2\int_0^\infty\!\!\!{\di r \di r'}J_{l+l'-n-n'+1}\left({eB\over2}rr'\right)(rr')^{l+l'-n-n'+2}e^{-eB(r^2+{r'}^2)/4}\nonumber\\
	&&\Bigg\{\prod_{x=r,r'}\Big[{|q_dB|}{F}_{nl,n'l'}(q_uB,|q_dB|;x)+{q_uB}{F}_{n'l',nl}(|q_dB|,q_uB;x)\Big]\nonumber\\
	&&+nn'{q_uB}{|q_dB|}\!\!\prod_{x=r,r'}\!\!\left({{F}_{nl,n'l'}(q_uB,|q_dB|;x)\over n'}\!+\!{{F}_{n'l',nl}(|q_dB|,q_uB;x)\over n}\right)\Bigg\},
	\end{eqnarray}
	the coefficient can be conveniently re-expressed as
	\begin{eqnarray}
	{\cal A}_{FL}&=&{-2N_ci\over S}\sum_{n=0}^\infty\sum_{l=0}\sum_{n'=0}^\infty\sum_{l'=0}\int_{-\infty}^{\infty}{dp_0\over2\pi}\int_{-\infty}^{\infty}{dp_z\over2\pi}
	{{n!n'!\over l!l'!}\left({q_uB\over 2}\right)^{l-n+1}\left({|q_dB|\over 2}\right)^{l'-n'+1}\over\left[\left({p}_0^{(l-n)+}\right)^2-(\varepsilon_n^u)^2\right]\left[\left({p}_0^{(n'-l')-}\right)^2-(\varepsilon_{n'}^d)^2\right]}\nonumber\\
	&&\Bigg\{H_{nl,n'l'}(q_uB,|q_dB|)+\left[{|q_dB|\over n'}G_{nl,n'l'}(q_uB,|q_dB|)+{q_uB\over n}G_{n'l',nl}(|q_dB|,q_uB)\right]\nonumber\\
	&&\ \ \times\left[\left({p}_0^{(l-n)+}\right)^2-(\varepsilon_n^u)^2+\left({p}_0^{(n'-l')-}\right)^2-(\varepsilon_{n'}^d)^2-\Omega_{nl,n'l'}^2\right]\Bigg\}.
	\end{eqnarray}
	Finally, by transferring to imaginary-time formalism and summing over the fermion Mastubara frequency, we get the finite temperature expression Eq.(\ref{AFL}).	
	
	To check the correctness of the above result, we consider the nonvanishing contributions from low Landau levels. First we study the contributions from $n=0,n'=1$ and $n=1,n'=0$. Utilizing the following results
	\begin{eqnarray}
	G_{0\,l,1\,l'}(q_uB,|q_dB|)&=&G_{0\,l',1\,l}(|q_dB|,q_uB)={(l+l')!\over eB}\left({2\over eB}\right)^{l+l'+1},\nonumber\\
	H_{0\,l,1\,l'}(q_uB,|q_dB|)&=&2{(l+l')!\over eB}\left({2\over eB}\right)^{l+l'+1} |q_dB|^2,\nonumber\\
	H_{1\,l,0\,l'}(q_uB,|q_dB|)&=&2{(l+l')!\over eB}\left({2\over eB}\right)^{l+l'+1} (q_uB)^2,
	\end{eqnarray}
	we obtain
	\begin{eqnarray}\label{A0C}
	{\cal A}_1&=&{-4N_ci\over S}\sum_{l=0}\sum_{l'=0}\int_{-\infty}^{\infty}{dp_0\over2\pi}\int_{-\infty}^{\infty}{dp_z\over2\pi}{(l+l')!\over l!l'!}\left({q_uB\over eB}\right)^{l+1}\left({|q_dB|\over eB}\right)^{l'+1}\nonumber\\
	&&\sum_{n+n'=1}{{p}_0^{(l-n)+}{p}_0^{(n'-l')-}-p_z^2-m^2\over\left[\left({p}_0^{(l-n)+}\right)^2-(\varepsilon_n^u)^2\right]\left[\left({p}_0^{(n'-l')-}\right)^2-(\varepsilon_{n'}^d)^2\right]}\nonumber\\
	&\stackrel{\Omega\rightarrow 0}{=}&{-2N_ci\over \pi}{q_uB|q_d|B\over eB}\int_{-\infty}^{\infty}{dp_0\over2\pi}\int_{-\infty}^{\infty}{dp_z\over2\pi}
	\left[{1\over{p}_0^2-(\varepsilon_1^d)^2}+{1\over{p}_0^2-(\varepsilon_1^u)^2}\right]\nonumber\\
	&=&-{N_c\over 2\pi}{q_uB|q_d|B\over eB}\int_{-\infty}^{\infty}{dp_z\over(2\pi)}\left[{1\over\varepsilon_1^d}+{1\over\varepsilon_1^u}\right].
	\end{eqnarray}
	Next we calculate the contribution from $n=n'=1$. Utilizing the following results
	\begin{eqnarray}
	G_{1\,l,1\,l'}(q_uB,|q_dB|)&=&{(l+l'-1)!\over (eB)^3}\left({2\over eB}\right)^{l+l'}\left[\left(|q_dB|l-q_uB l'\right)^2-(l+l')(q_uB)^2\right],\nonumber\\
	H_{1\,l,1\,l'}(q_uB,|q_dB|)&=&2|q_dB|eBG_{1\,l,1\,l'}(q_uB,|q_dB|)+2q_uBeBG_{1\,l',1\,l}(|q_dB|,q_uB)\nonumber\\
	&&-8|q_dB|q_uB{(l+l'-1)!\over (eB)^3}\left({2\over eB}\right)^{l+l'}\left[\left(|q_dB|l-q_uB l'\right)^2+(l+l')|q_dB|q_uB\right],
	\end{eqnarray}
	we obtain
	\begin{eqnarray}\label{A1C}
	{\cal A}_2
	&=&{-4N_ci\over S}\sum_{l=0}\sum_{l'=0}\int_{-\infty}^{\infty}{dp_0\over2\pi}\int_{-\infty}^{\infty}{dp_z\over2\pi}
	{(l+l'-1)!\left({q_u\over e}\right)^{l}\left({|q_d|\over e}\right)^{l'}\over l!l'!\left[\left({p}_0^{(l-1)+}\right)^2-(\varepsilon_1^u)^2\right]\left[\left({p}_0^{(1-l')-}\right)^2-(\varepsilon_{1}^d)^2\right]}\nonumber\\
	&&\Bigg\{\Big({p}_0^{(l-1)+}{p}_0^{(1-l')-}-p_z^2-m^2-4eB{q_u\over e}{|q_d|\over e}\Big)\left({|q_d|\over e}l-{q_u\over e}l'\right)^2\nonumber\\
	&&\ \ -\left({p}_0^{(l-1)+}{p}_0^{(1-l')-}-p_z^2-m^2+4eB{q_u\over e}{|q_d|\over e}\right)(l+l')\left({|q_d|\over e}\right)\left({q_u\over e}\right)\Bigg\}\nonumber\\
	&\stackrel{\Omega\rightarrow 0}{=}&{-2N_ci\over \pi}{q_uB|q_d|B\over eB}\int_{-\infty}^{\infty}{dp_0\over2\pi}\int_{-\infty}^{\infty}{dp_z\over2\pi}
	{1\over\left[{p}_0^2-(\varepsilon_1^u)^2\right]\left[{p}_0^2-(\varepsilon_{1}^d)^2\right]}\left\{\left({p}_0^2-p_z^2-m^2-4eB{q_u\over e}{|q_d|\over e}\right)\right.\nonumber\\
	&&\left.-\left({p}_0^2-p_z^2-m^2+4eB{q_u\over e}{|q_d|\over e}\right)\right\}\nonumber\\
	&=&-{16N_c\over \pi}(eB)^2\left({q_u\over e}{|q_d|\over e}\right)^2\int_{-\infty}^{\infty}{dp_4\over2\pi}\int_{-\infty}^{\infty}{dp_z\over2\pi}{1\over\left[{p}_4^2+(\varepsilon_1^u)^2\right]\left[{p}_4^2+(\varepsilon_{1}^d)^2\right]}.
	\end{eqnarray}
	In the vanishing rotation limit ($\Omega=0$), we can compare the above results  with the those obtained by using the effective fermion propagator in Euclidean space~\cite{Miransky:2015ava}. With the Euclidean method, we obtain
	\begin{eqnarray}
	{\cal A}_1&=&-\int{d^4p\over{(2\pi)^4}}{e^{-(p_x^2+p_y^2)[1/(q_uB)+1/(|q_dB|)]}\over \left[p_4^2+\left(\varepsilon_1^u\right)^2\right]\left[p_4^2+\left(\varepsilon_0^d\right)^2\right]}{\rm Tr}\Bigg\{(m-p_4\gamma_4-p_z\gamma^3)\left[(1+i\gamma^1\gamma^2)L_1\left({2(p_x^2+p_y^2)\over q_uB}\right)-(1-i\gamma^1\gamma^2)\right]\nonumber\\
	&&(-m-p_4\gamma_4-p_z\gamma^3)(1-i\gamma^1\gamma^2)\Bigg\}-\int{d^4p\over{(2\pi)^4}}{e^{-(p_x^2+p_y^2)[1/(q_uB)+1/(|q_dB|)]}\over \left[p_4^2+\left(\varepsilon_0^u\right)^2\right]\left[p_4^2+\left(\varepsilon_1^d\right)^2\right]}{\rm Tr}\Bigg\{(m-p_4\gamma_4-p_z\gamma^3)(1+i\gamma^1\gamma^2)\nonumber\\
	&&(-m-p_4\gamma_4-p_z\gamma^3)\left[(1-i\gamma^1\gamma^2)L_1\left({2(p_x^2+p_y^2)\over |q_dB|}\right)-(1+i\gamma^1\gamma^2)\right]\Bigg\}\nonumber\\
	&=&-8N_c\int{d^4p\over{(2\pi)^4}}e^{-(p_x^2+p_y^2)[1/(q_uB)+1/(|q_dB|)]}\left[{1\over p_4^2\!+\!\left(\varepsilon_1^u\right)^2}\!+\!{1\over p_4^2\!+\!\left(\varepsilon_1^d\right)^2}\right]\nonumber\\
	&=&-{N_c\over 2\pi}{q_uB|q_d|B\over eB}\int_{-\infty}^{\infty}{dp_z\over(2\pi)}\left[{1\over\varepsilon_1^d}\!+\!{1\over\varepsilon_1^u}\right]\label{A0M}
	\end{eqnarray}
	and
	\begin{eqnarray}
	{\cal A}_2&=&\int{d^4p\over{(2\pi)^4}}{e^{-(p_x^2+p_y^2)[1/(q_uB)+1/(|q_dB|)]}\over \left[p_4^2+\left(\varepsilon_1^u\right)^2\right]\left[p_4^2+\left(\varepsilon_1^d\right)^2\right]}{\rm Tr}\Bigg\{(m-p_4\gamma_4-p_z\gamma^3)\left[(1+i\gamma^1\gamma^2)L_1\left({2(p_x^2+p_y^2)\over q_uB}\right)-(1-i\gamma^1\gamma^2)\right]\nonumber\\
	&&+4(p_x\gamma^1+p_y\gamma^2)\Bigg\}\Bigg\{(-m-p_4\gamma_4-p_z\gamma^3)\left[(1-i\gamma^1\gamma^2)L_1\left({2(p_x^2+p_y^2)\over |q_dB|}\right)-(1+i\gamma^1\gamma^2)\right]+4(p_x\gamma^1+p_y\gamma^2)\Bigg\}\nonumber\\
	&=&8\int{d^4p\over{(2\pi)^4}}{e^{-(p_x^2+p_y^2)[1/(q_uB)+1/(|q_dB|)]}\over \left[p_4^2+\left(\varepsilon_1^u\right)^2\right]\left[p_4^2+\left(\varepsilon_1^d\right)^2\right]}\left\{(p_4^2+p_z^2+m^2)\left[L_1\left({2(p_x^2+p_y^2)\over q_uB}\right)+L_1\left({2(p_x^2+p_y^2)\over |q_dB|}\right)\right]-8(p_x^2+p_y^2)\right\}\nonumber\\
	&=&-{16N_c\over \pi}(eB)^2\left({q_uB\over eB}{|q_dB|\over eB}\right)^2\int_{-\infty}^{\infty}{dp_4\over2\pi}\int_{-\infty}^{\infty}{dp_z\over2\pi}{1\over \left[p_4^2+\left(\varepsilon_1^u\right)^2\right]\left[p_4^2+\left(\varepsilon_1^d\right)^2\right]}.\label{A1M}
	\end{eqnarray}
	Thus, for the two types of contributions, the results from two different methods are consistent with each other in the vanishing rotation limit.
\end{widetext}

\end{document}